\documentclass[12pt,preprint]{aastex}

\usepackage{rotating}

\slugcomment{Accepted for publication in Apj}
\shorttitle{Microlensing mass function}
\shortauthors{Mediavilla et al.}

\begin{document}

%\title{Applications of the Statistics of Caustic Crossings to the Study of Microlensing in Q~2237+0305}

\title{Statistics of Microlensing Caustic Crossings in Q~2237+0305: Peculiar Velocity of the Lens Galaxy and Accretion Disk Size}

%\author{MEDIAVILLA et AL.}

\author{E. MEDIAVILLA\altaffilmark{1,2}, J. JIMENEZ-VICENTE\altaffilmark{3,4}, J. A. MU\~NOZ\altaffilmark{5}, T. MEDIAVILLA\altaffilmark{6}, O. ARIZA\altaffilmark{6}}

\altaffiltext{1}{Instituto de Astrof\'{\i}sica de Canarias, V\'{\i}a L\'actea S/N, La Laguna 38200, Tenerife, Spain}
\altaffiltext{2}{Departamento de Astrof\'{\i}sica, Universidad de la Laguna, La Laguna 38200, Tenerife, Spain}
\altaffiltext{3}{Departamento de F\'{\i}sica Te\'orica y del Cosmos, Universidad de Granada, Campus de Fuentenueva, 18071 Granada, Spain}
\altaffiltext{4}{Instituto Carlos I de F\'{\i}sica Te\'orica y Computacional, Universidad de Granada, 18071 Granada, Spain}
\altaffiltext{5}{Departamento de Astronom\'{\i}a y Astrof\'{\i}sica, Universidad de Valencia, 46100 Burjassot, Valencia, Spain.}
\altaffiltext{6}{Departamento de Estad\'{\i}stica e Investigaci\'on Operativa, Universidad de C\'adiz, Avda Ram\'on Puyol s/n, 11202, Algeciras, C\'adiz, Spain}

\begin{abstract}

We use the statistics of caustic crossings induced by microlensing in the lens
system Q~2237+0305 to study the lens galaxy peculiar velocity. We calculate the
caustic crossing rates for a comprehensive family of stellar mass functions and
find a dependence of the average number of caustic crossings with the effective
transverse velocity and the average mass,  $\langle n \rangle  \propto {v_{eff}
/ \sqrt{\langle m \rangle}}$,  equivalent to the theoretical prediction for the
case of microlenses with identical masses. We explore the possibilities of the method to measure $v_{eff}$ using
the $\sim$12 years of OGLE monitoring of the four images of Q 2237+0305. To
determine a lower limit for $v_{eff}$ we count, conservatively, a single caustic
crossing for each one of the 4 high magnification events identified in the
literature (plus one additional proposed by us) obtaining $v_{eff} \gtrsim
240\sqrt{\langle m \rangle/0.17M_\odot}\rm\,km\, s^{-1}$ at 68\% of confidence.
From this value and the average $FWHM$ of 4 high magnification events we obtain
a lower limit of $r_s \gtrsim 1.4 \sqrt{\langle m \rangle/0.17M_\odot}$
light-days for the  radius of the source ($r_s=FWHM/2.35$). Tentative
identification of 3 additional caustic crossing events leads to estimates of
$v_{eff}\simeq (493\pm 246)\sqrt{\langle m \rangle/0.17M_\odot}\rm\,km\, s^{-1}$
for the effective transverse velocity and of $r_s \simeq (2.7\pm
1.3)\sqrt{\langle m \rangle/0.17M_\odot}$ light-days for the source size. The
estimated transverse peculiar velocity of the galaxy is $v_t 
\simeq(429\pm246)\sqrt{\langle m \rangle/0.17M_\odot}\rm\,km\, s^{-1}$.
\end{abstract}


\section{Introduction}

The mean (or macro) magnification of any of the images of a gravitationally lensed quasar is calculated supposing that the matter in the lens galaxy follows a smooth distribution. On parsec scales, however, the distribution of normal (not-dark) matter is highly inhomogeneous, concentrated in almost point-like stars surrounded by vast regions of emptiness. The granulation of a part of the galaxy mass into stars, while microscopic in comparison with the galaxy dimensions, induces strong local changes in the magnification (quasar microlensing, Chang \& Refsdal 1979, 1984; see also Kochanek 2004 and Wambsganss 2006). The scenario of quasar microlensing involves interesting information related to the lens galaxy (e.g. the stellar mass function or the peculiar velocity of the lens galaxy) and to the quasar source (e.g. the continuum source size).

The method commonly used to extract this information is based in the statistical comparison of observed light curves of lensed quasar images with light curves simulated for
different values of the physical parameters of interest (source size, peculiar velocity of the lens galaxy, mean mass of the stars, slope of the mass function, etc.). This statistical analysis, light curve fitting, makes use of all the information in the light curves but faces several problems. First, the most conspicuous
features associated to microlensing are caustic crossings whose statistics are related to the relative transverse velocity between quasar and galaxy and may be also sensitive, by an amount that we will discuss later, to the microlens mass function (see, e.g., 
Wyithe \& Turner 2001, Congdon et al. 2007). However, in optical observations caustic crossings may appear smoothed and blended due to the non-negligible size of the quasar source (see e.g. Jim{\'e}nez-Vicente et al. 2012 and references therein). Second, there is a degeneracy involving source size and the transverse velocity of the lens galaxy (see, e.g., Wyithe et al.  2000b). Third, the baseline for no microlensing magnification (i.e. the macro model magnification) is often unknown and/or affected by intrinsic variability, differential extinction, undetected substructure, or contamination from the host galaxy of the quasar. For these reasons an adjustable offset in magnitudes is typically used to fit the light curves (see e.g. Kochanek 2004). Finally, even with this adjustable offset, a huge computational effort is needed to obtain good fits of long light curves (sometimes the curves have to be split into smaller pieces, see e.g. Poindexter \& Kochanek 2010).

A possible solution to minimizing these degeneracies would be to use X-Ray emission arising from a source small enough so that each single caustic crossing appears as an isolated high magnification event. In this case, we could measure a caustic crossing rate that does not depend on the source size, breaking the size/velocity degeneracy. For example, the limiting but interesting case of 0 caustic crossings was used by Gil-Merino et al. (2005) to constrain the transverse velocity in Q 2237+0305 (see also Wyithe et al. 1999). Once the velocity is known, we can examine the individual high magnification events to constrain the size and, eventually, to study the luminosity profile of the source. Note that the caustic crossing rate is independent of intrinsic source variability, extinction or any other source of flux contamination. In a subsequent step, the estimates for the galaxy velocity and/or the quasar size may be used as priors to make easier the application of the thorough method of the light curve fitting that uses not only the information involved in the caustic crossings but all the information contained in the light curves. 

Unfortunately, the available X-Ray light curves (Chen et al. 2011) are irregularly sampled and cover relatively short periods of time. The best light curves we have at our disposal are the OGLE optical light curves from Q 2237+0305 (Wozniak et al. 2000, Udalski et al. 2006) that cover a long period of time with a regular sampling albeit with the effects of a significant source size. This problem seems to be surmountable since high magnification caustic crossing events have been identified in studies based on the Q 2237+0305 optical light curves (Wyithe et al. 2000c, Shalyapin 2001, Goicoechea et al. 2003, Kochanek 2004, Gil-Merino et al. 2006, Anguita et al. 2008, Eigenbrod et al. 2008a,b, Mosquera et al. 2009, Abolmasov \& Shakura 2012, Mosquera et al. 2013). These studies are, in general, focused on a small fraction of the light curve around one isolated high magnification event. A global analysis of the available light curves of the 4 lensed images of Q 2237+0305 could consistently study the impact of caustic blending by comparing the FWHM of the events with the mean separation between caustics.

The other question that we should also consider in the analysis is how the microlenses mass function affects the estimate of the caustic crossing rate. Previous works suggest that the rate of microlensing high magnification events does not depend on the details of the mass function but rather on the mean microlens mass (e.g. Witt, Kayser \& Refsdal, 1993; Wyithe et al. 2000a). However these studies are limited to very specific mass functions (basically the Salpeter law, Salpeter 1955) with relatively high mean ($\geq 0.3\ M_\odot$) and minimum ($0.1\ M_\odot$) masses \footnote{Other studies based on microlensing magnification statistics also indicate that the distribution of magnifications is not very sensitive to the microlenses mass function (Wyithe \& Turner 2001) although this could be true only for small sized sources (Congdon et al. 2007).}.  To explore the sensitivity of microlensing to the abundance of low mass stars we will analyze the impact in the caustic crossing rate of a comprehensive family of mass functions spanning a wide range of slopes at the low mass end. This will complicate the analysis but it is worth to explore the role of low mass ($0.1\ M_\odot \geq M\geq  0.006\ M_\odot$) microlenses, especially taken into account that the shape and universality of the substellar mass function are still rather uncertain (see e.g. Bastian et al. 2010).

The first objective of the present work is, then, to analyze the impact of the transverse effective velocity and of the microlenses mass function in the statistics of caustic crossings.  In a second step, we will jointly study the four Q 2237+0305 optical light curves and use the caustic crossing statistics to estimate the transverse velocity and to constrain the quasar source size from the individual high magnifications events. We will use both velocity and size estimates to discuss the consistency of the procedure and the impact of undetected caustic crossings in the estimates.  According to this approach, we will start in \S 2 by describing our parametrization of the mass function and the statistical method based in caustic crossing
counts.  In \S 3 we explore the application of this analysis to the available optical and future X-Ray observations of Q~2237+0305. Finally, the main conclusions are
presented in \S 4.

\section{Caustic Crossing Statistical Analysis \label{DA}}

\subsection{Identical Mass Microlenses}

For use as reference in the next section, we first review the case in which all the microlenses have identical mass, $m$ (see e.g. Wyithe et al. 2000a). If we take into account that the average number of caustic crossings, $\langle n \rangle$, in a track of length $l$ is proportional to the number of Einstein radii along the track, and that an Einstein radius is proportional to $\sqrt{m}$, we have:
$\langle n \rangle  \propto {l /  \sqrt{m}}$. On the other hand, the effective transverse velocity of the caustics, $v_{eff}$, is related     to the track length through the time elapsed to travel the track, $t$, $ v_{eff}={l/t}$. Thus, for a given number of crossings, $n$, observed in a time $t$, there is a proportionality between $v_{eff}$ and $\sqrt{m}$,

\begin{equation}
{v_{eff} \over  \sqrt{m}}\propto \langle n \rangle.
\end{equation}

\subsection{Microlenses Distributed According to a Parametric Stellar Mass Function}

\begin{figure}[h]
%\plotone{/home/emg/polygon_2011/programas/FIG1.eps}
\plotone{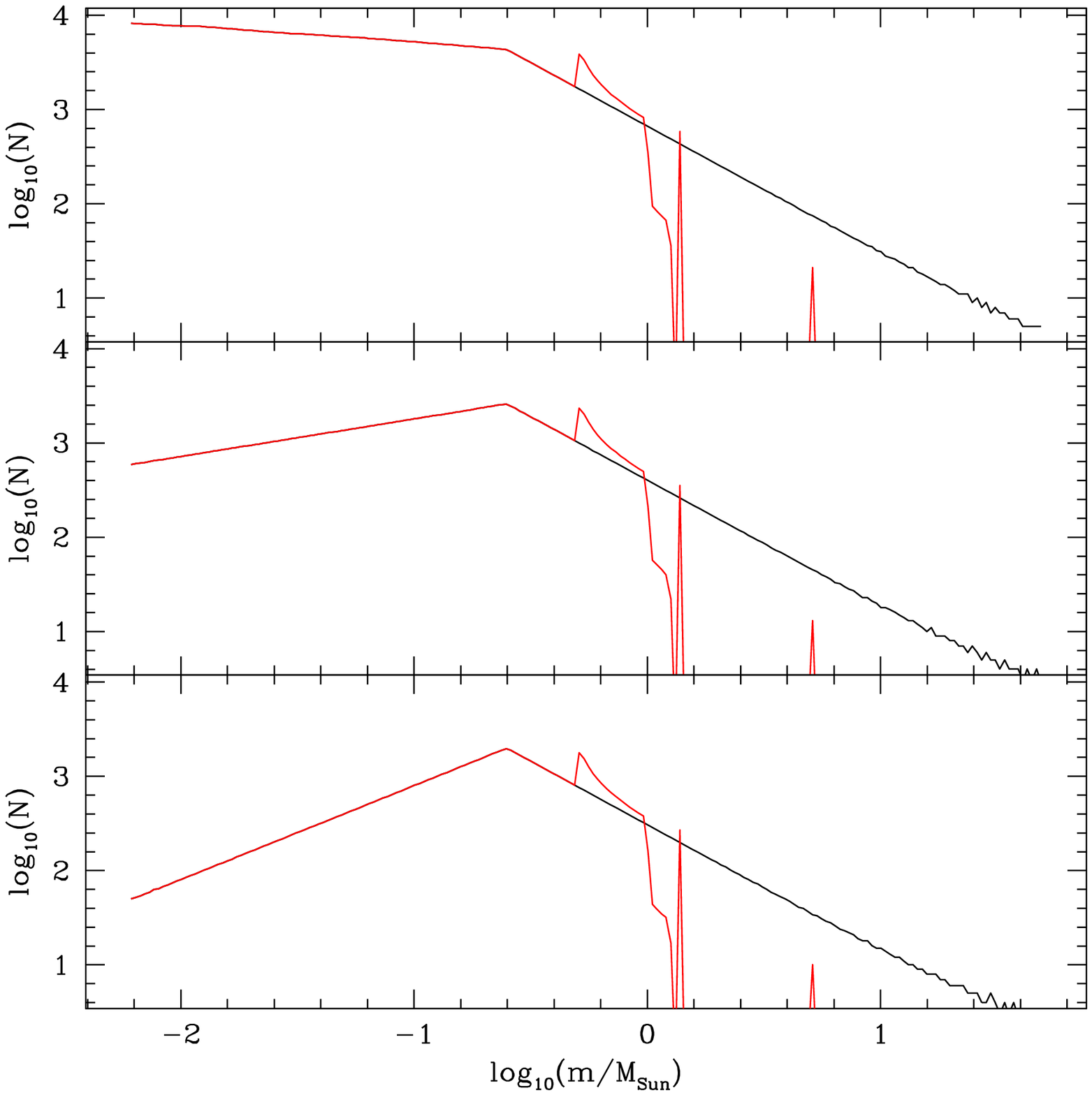}
\caption{Present day (red) and initial (black) mass functions for $\alpha_1=$ 0 (up), 0.6 (middle), and 1.2 (bottom). See text.  \label{pdmf}}
\end{figure}

We will start defining the stellar mass function. A double power-law (Kroupa et al. 1993), $\Delta N / \Delta m \propto m^{-\alpha}$, with slopes $\alpha_1$ and $\alpha_2$ in
the $0.006 M_\odot\le m < 0.25 M_\odot$ and $0.25 M_\odot\le m \le 50 M_\odot$ ranges, respectively, will be used to describe
the initial mass function (IMF). We adopt the $0.006 M_\odot$ lower limit from Pe\~na-Ram\'\i rez et al. (2012) and the characteristic
mass that separates the two power-laws, $0.25 M_\odot$,  from Bastian et al. (2010; see their Figure 3). We adopt a $50
M_\odot$ maximum mass. Following the general consensus (Bastian et al.
2010) we will take a Salpeter law (Salpeter 1955) for the high-mass power law, $\alpha_2=-2.35$. On the contrary,
$\alpha_1$ is rather uncertain (and perhaps not universal) showing a range of variation from $\sim$0 to 1.2 (see Figure 2 from
Bastian et al. 2010). Thus, we will consider by now $\alpha_1$ as the free parameter. With this parametrization of the mass
function, we can probe the sensitivity of microlensing to the presence of low-mass stars. From this IMF we derive the present day mass function (PDMF) following the steps described in Poindexter \& Kochanek (2010) to model the stars with $m > 1 M_\odot$ as remnants (either White Dwarfs, Neutron Stars or Black Holes). In Figure \ref{pdmf} we show (for an arbitrarily large number of stars) the IMF and PDMF for $\alpha_1=0, 0.6, 1.2$. To include in our study even more extreme cases, we will consider $\alpha_1$ in a range of variation from $-$0.2 to 1.4. Note that for the adopted parametric family of stellar mass functions, there is a one to one relationship between the slope at the low mass end, $\alpha_1$, and the average mass, $\langle m \rangle$. We will make use of this relationship to compare with the case of identical mass microlenses.

\begin{figure}[h]
%\plotone{newf2.ps}
\plotone{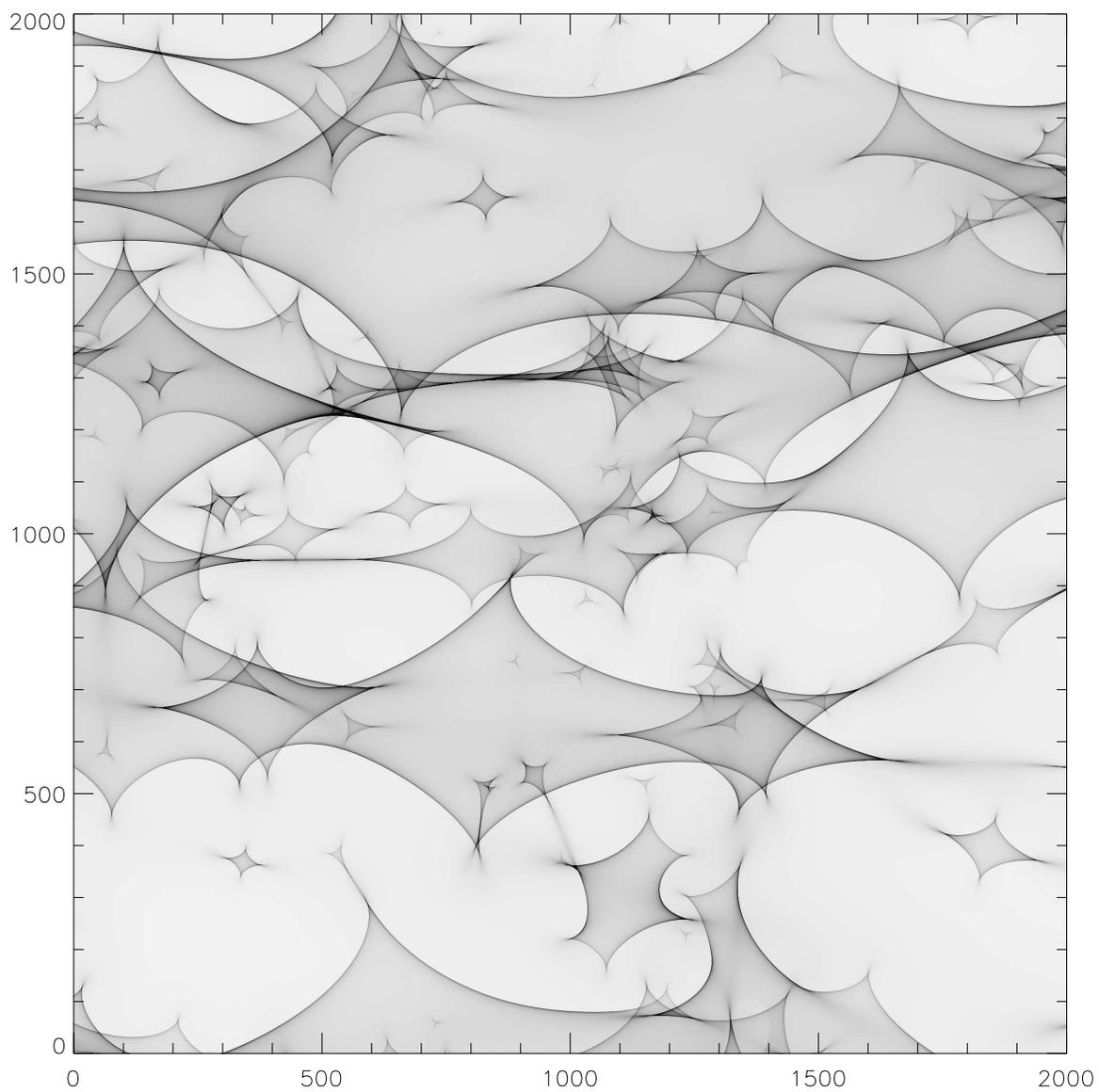}
\caption{Example of magnification map for image D of Q~2237+0305. \label{map}}
\end{figure}

\begin{figure}[h]
%\plotone{f2ev2.ps}
\plotone{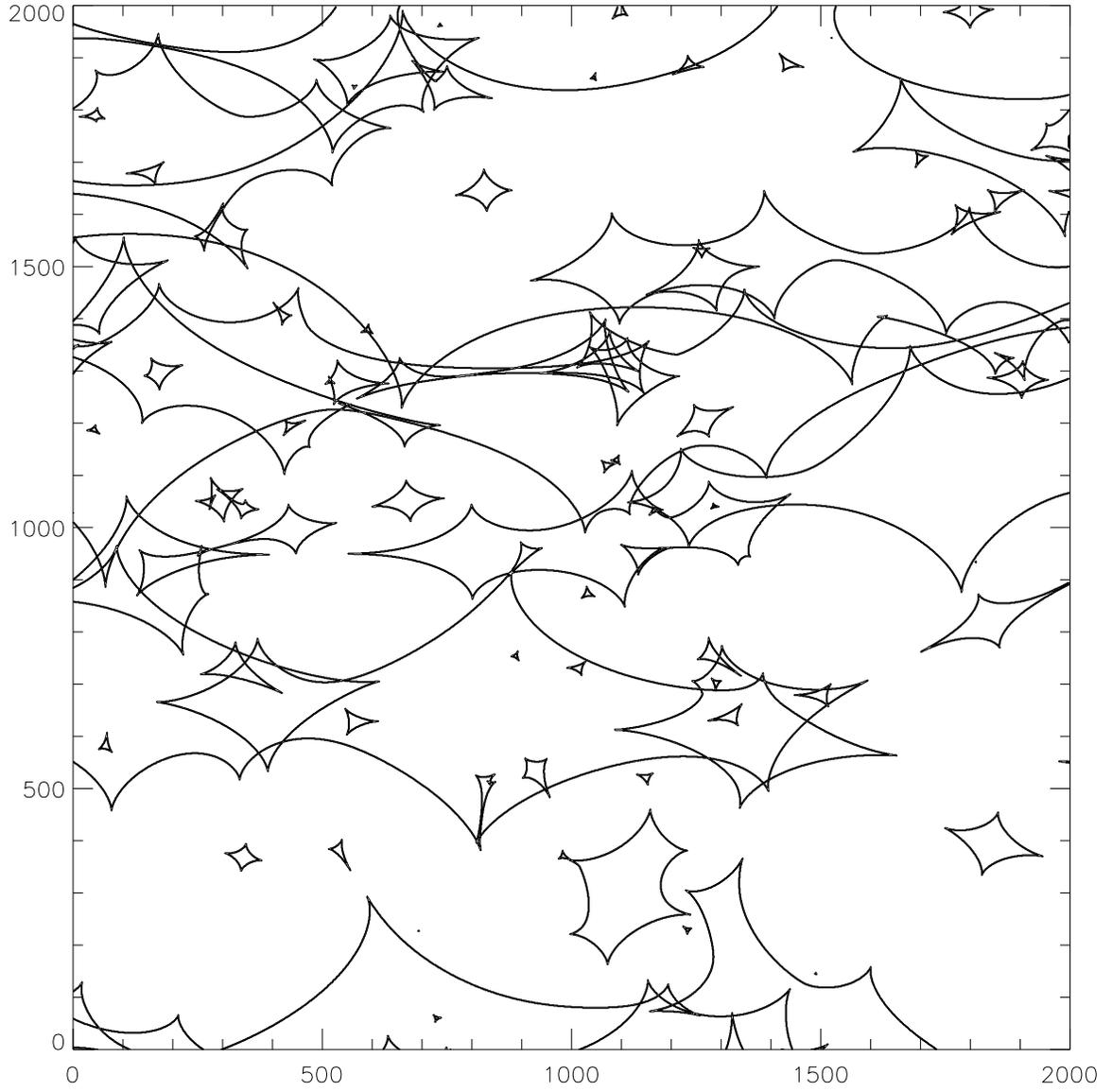}
\caption{Map of caustics corresponding to the magnification map of Figure \ref{map}. \label{causticas}}
\end{figure}

%{In principle, we could have left free another parameter of interest (like the lower mass cutoff), however the degeneracy with the peculiar transverse velocity of the lens galaxy would make neccesary to fit three parameters and this will complicate very much the practical aplication of the method.  A two parameters (low-mass slope and lower mass cutoff) fit to the mass function would be reasonable if the transverse velocity were obtained from other means.}
Our purpose is to know how the coexistence of microlenses of different masses distributed according to the parametric PDMF defined above and the variation of the proportion between stars of different masses (that is the variation of the lower end slope of the PDMF, $\alpha_1$) change the theoretical results for identical mass particles presented in the previous section%\footnote{By analogy with the identical mass microlenses case, we will take the square root of the mean microlens mass, $\sqrt{\langle m \rangle}$, as the physical parameter representing the mass function.} 
. To this end, we perform an statistical analysis based on the number of caustic crossings, $n$, along different tracks (randomly
selected) on the source plane. In studies based on light-curve fitting, it is usual to use a magnification map to simulate microlensing induced variability along the
tracks. In our case we will use the Inverse Polygon Mapping (Mediavilla et al. 2006, 2011) to directly obtain maps of caustic curves (following the technique described in Mediavilla et al. 2011) that can be straightforwardly used to count the number of caustic crossings along each track.  We have calculated magnification and caustic maps for the four images of Q~2237+0305 taking the following values for the convergence and shear: $\kappa_A=0.36$, $\gamma_A=0.40$, $\kappa_B=0.36$, $\gamma_B=0.42$, $\kappa_C=0.69$, $\gamma_C=0.71$, $\kappa_D=0.59$, $\gamma_D=0.61$ {(Schmidt et al. 1998)}. We will suppose that the fraction of mass in stars is 100\%. The maps have $2000 \times 2000$ pixels of 0.2 light-days each one.  We have considered 9 different stellar mass functions with values of the low-mass power-law slope, $\alpha_1=-0.2,0,0.2,0.4,0.6,0.8,1.0,1.2,1.4$.  In Figures \ref{map} and \ref{causticas} we show an example of magnification map and its corresponding map of caustics for image D. 

From the maps of caustics we  have calculated the probability of the number of crossings, $p(n|l,\alpha_1)$, conditioned
to $\alpha_1$ and to the length of the track, $l$. Taking into account that for our PDMFs there is a one to one correspondence between $\alpha_1$ and $\langle m \rangle$ we have also trivially calculated $p(n|l,\sqrt{\langle m \rangle})$ (note that the sampling of PDFMs is linear in $\alpha_1$ not in $\sqrt{\langle m \rangle}$). To improve the statistics, we have used 10 maps generated with different random seeds to obtain each likelihood function, $p(n|l,\sqrt{\langle m \rangle})$. For the tracks we have considered the following lengths, $l=2\cdot 10^{n/10}$ light-days with $n=0,1,...,25$ $(2\leqslant l \leqslant 632.5)$. The longest tracks are larger than the side of the maps. To deal with these tracks we have constructed mosaics of 3 X 3 maps.

\begin{figure}[h]
%\epsscale{0.8}
\plotone{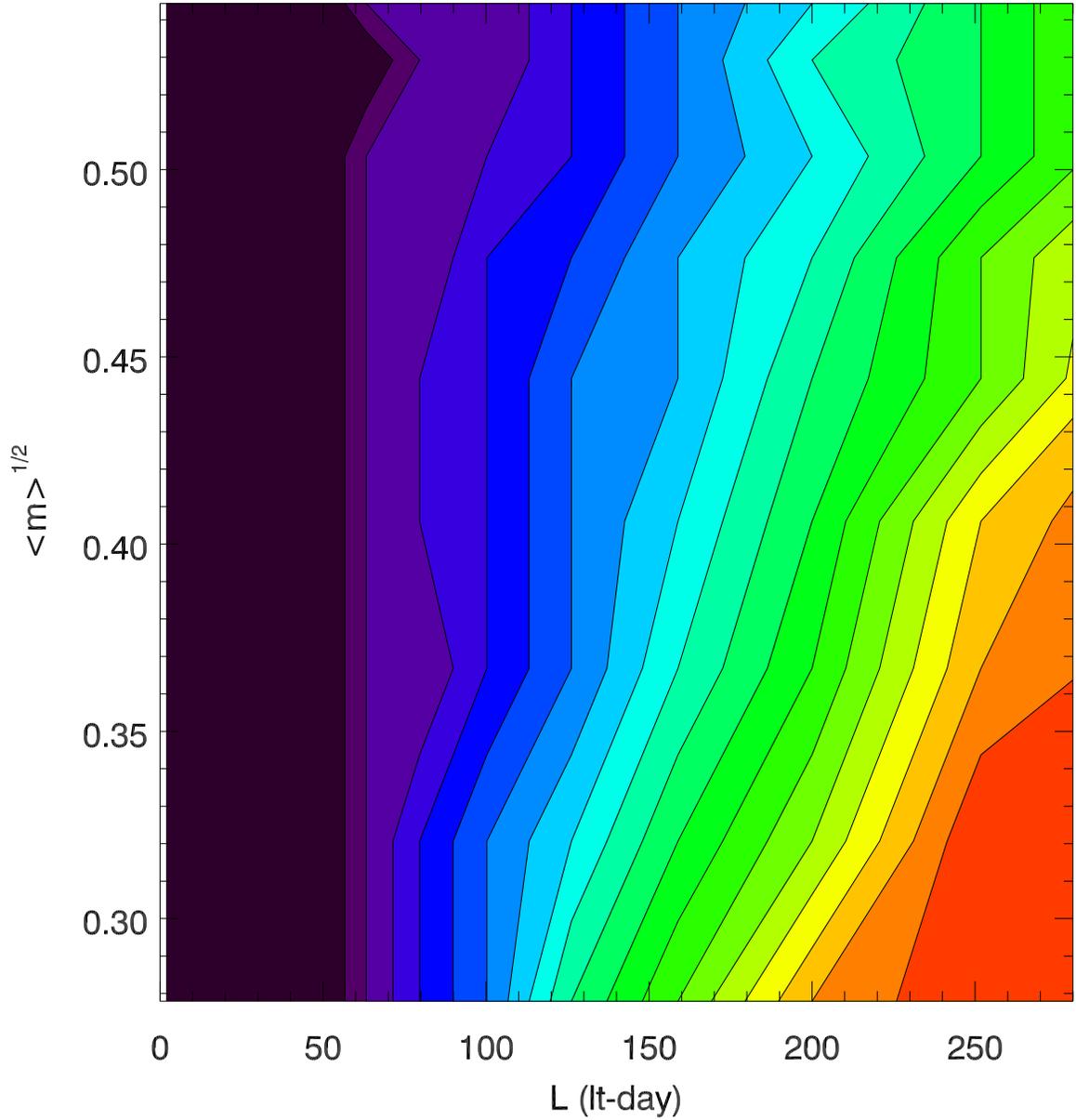}
\caption{Most likely number of crossings,  $n_{peak}(l,\sqrt{\langle m\rangle})$, for each track length, $l$, and square root of the average mass of the present day mass function, $\sqrt{\langle m\rangle}$. Contours range from 0 (black) to 19 crossings (red) in steps of 1 crossing. See text.  \label{npeak}}
\end{figure}

From the likelihoods, $p(n|l,\sqrt{\langle m \rangle})$, we have calculated (Figure \ref{npeak}) the most probable number of crossings for each value of $l$ and $\sqrt{\langle m \rangle}$, $n_{peak}(l,\sqrt{\langle m \rangle})$. In this Figure we can appreciate a clear covariance between $l$ and $\sqrt{\langle m \rangle}$ similar to that found in the case of identical mass microlenses. On the other hand, the change in $n_{peak}$ is much more sensitive to the track length than to the mean mass of the PDMF.

\begin{figure}[h]
\epsscale{0.8}
\plotone{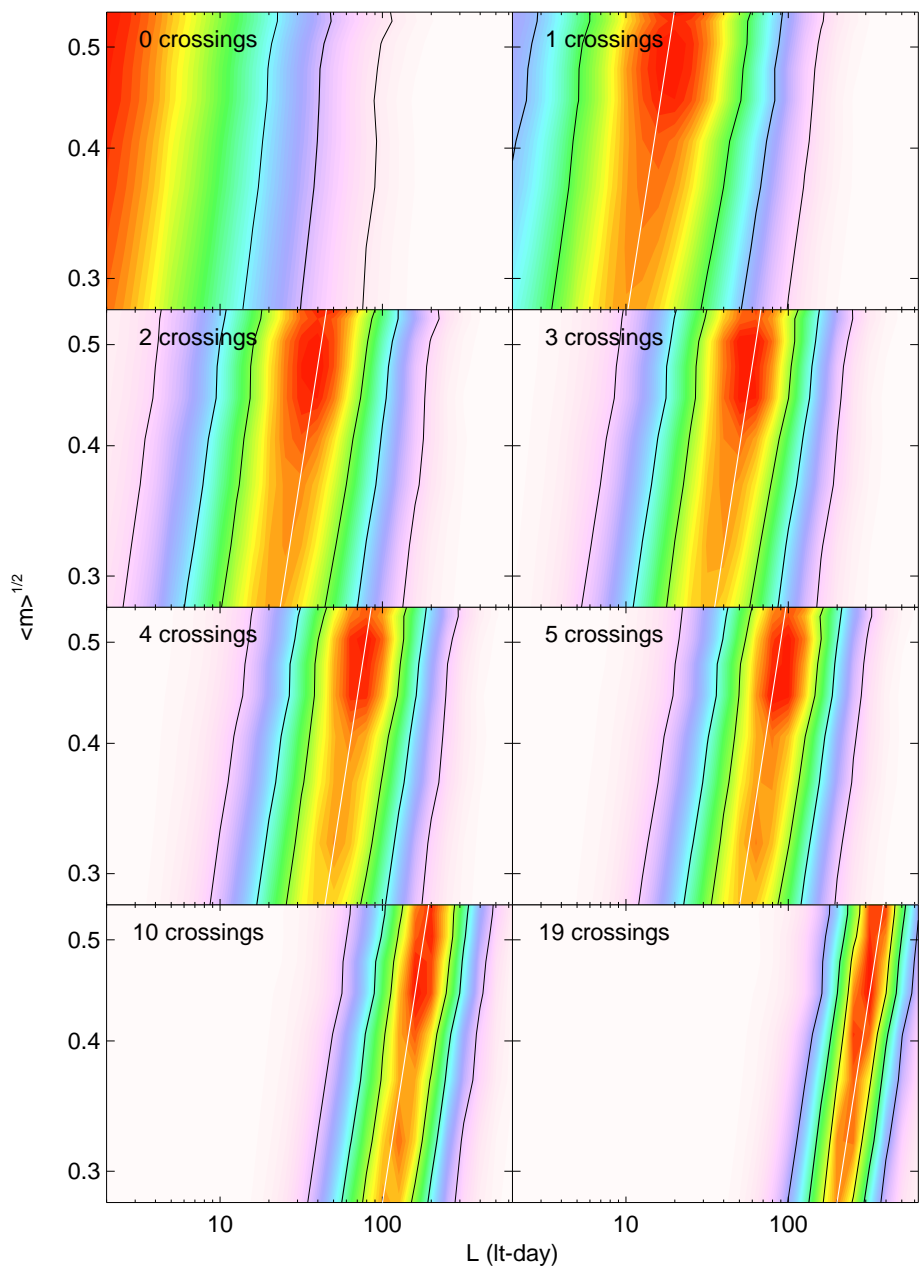}
\caption{2D probability density functions, $p(l,\sqrt{\langle m\rangle}|n)$, of the track length, $l$, and the square root of the average mass of the present day mass function, conditioned to different values of the number of crossings, $n$. The white line correspond to the predictions of the $\langle n \rangle  \propto {l /  \sqrt{\langle m\rangle}}$ law (see text). Contours are spaced by 0.5 sigma intervals. \label{pdfs}}
\end{figure}

\begin{figure}[h]
\plotone{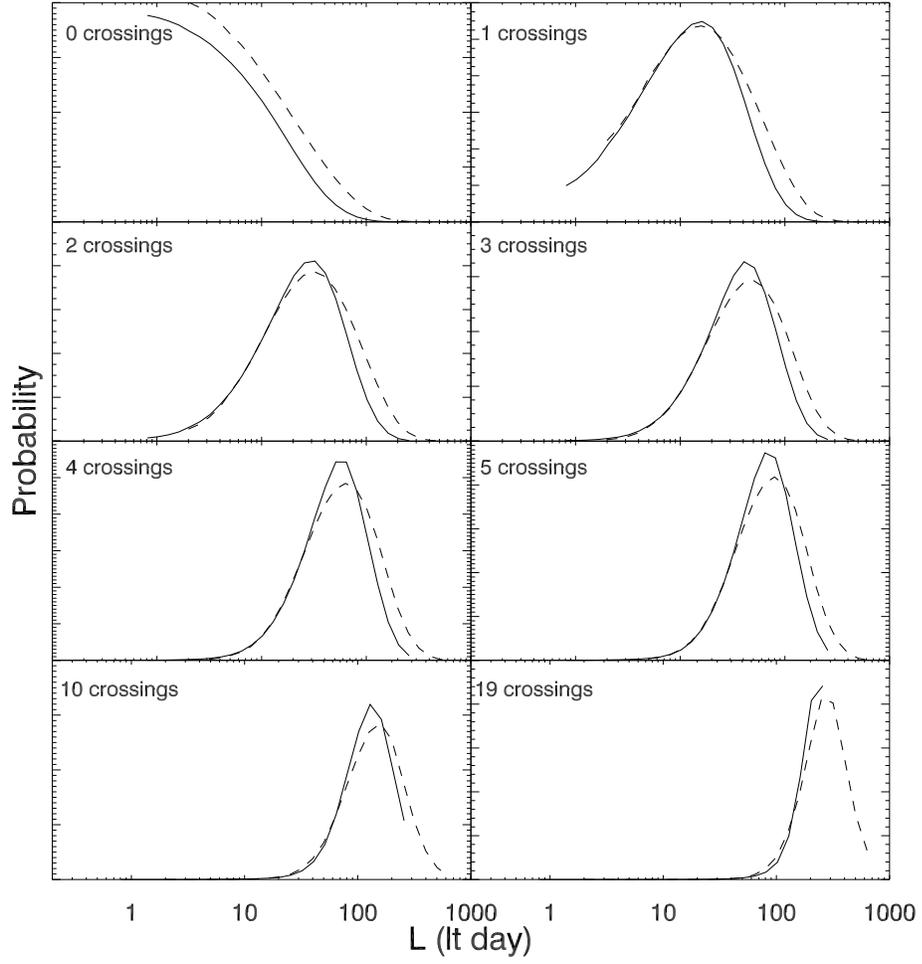}
\caption{Dashed lines: slices of the 2D probability density functions in Figure \ref{pdfs} for $\sqrt{<m>}= 0.4$, $p(l,\sqrt{\langle m\rangle}= 0.4|n)$. Continuous lines: same slices for the case of identical mass particles, $p(l,\sqrt{m}= 0.4|n)$. See text. \label{comp}}
\end{figure}

Using Bayes theorem we can also obtain the probability of $l$ and $\sqrt{\langle m \rangle}$ for a given number of crossings. We have used a logarithmic (linear) prior on $l$ ($\alpha_1$). In Figure \ref{pdfs} we plot for image D the probabilities, $p(l,\sqrt{\langle m \rangle}|n)$,  for $n=0,1,2,3,4,5,10,19$.   The main result of these Figures is, once more, the strong covariance between $\sqrt{\langle m \rangle}$ and $l$. In fact the ridges of the pdfs are very well matched by

\begin{equation}
\label{track}
{l \over \langle n \rangle \sqrt{\langle m \rangle/M_\odot}}= l_1, 
\end{equation}
where $l_1$ is a constant that can be interpreted as the average track length to obtain one caustic crossing for microlenses of $1M_\odot$. That means that for distributions in which stars with different masses coexist (in the wide range of ratios between small and large masses considered here) the maximum probability relationship between $\sqrt{\langle m \rangle}$ and $l$ is equivalent to the one deduced in the case of identical masses. In Figure \ref{comp} we plot $p(l,\sqrt{\langle m \rangle}= 0.4 M_\odot|n)$  and the corresponding probability distributions for the case of identical mass particles $p(l,\sqrt{m }=0.4 M_\odot|n)$. Some slight differences between peaks, centroids and FWHMs can be appreciated and, in fact, in Figure \ref{pdfs} there are maxima of probability that break the $l \propto \sqrt{\langle m \rangle}$ degeneracy but only at a very low level of significance in $\sqrt{\langle m \rangle}$. On the other hand, the wide range of values of $\alpha_1$ explored, makes us think that  Eq. \ref{track} is very general.

Thus, according to our simulations the degeneracy intrinsical to a distribution of identical mass microlenses extends to the case of unequal particles, ${v_{eff} /  \sqrt{\langle m \rangle}}\propto \langle n \rangle$, not only for Salpeter law with a lower cutoff of $0.1 M_\odot$  (Witt, Kayser \& Refsdal, 1993; Wyithe et al. 2000a) but also for the wide range of values for $\alpha_1$ and the $0.006 M_\odot$ minimum mass considered here.

In principle, it may seem that this degeneracy strongly limits the usefulness of the caustic crossing counts to study either the transverse velocity or the mass function. However,  when $\alpha_1$  ranges from $-$0.2 to 1.4, the PDMF mean mass, $\langle m \rangle$, ranges from 0.30 to 0.08$M_\odot$ and the parameter of interest, $\sqrt{\langle m \rangle}$, from 0.54 to 0.28$\sqrt{M_\odot}$. Thus, the estimate of the peculiar velocity based on the number of caustic crossings changes by only a 50\% over the whole range of $\sqrt{\langle m \rangle}$. This is a relatively modest uncertainty taking into account that we have considered extreme cases for the slope at the low mass end of the IMF, $\alpha_1$. 

Then, the low sensitivity of the number of caustic crossings to the expected range of variation in the average stellar mass reduces the usefulness of $ \langle n \rangle$ to study the stellar mass function. For the same reason, we can have reasonable good expectations to estimate the transverse effective velocity.

\section{Constraints on the Transverse Effective Velocity in Q~2237+0305}

\begin{figure}[h]
%\plotone{/home/emg/2237/ogle/over_ABCD.eps}
\plotone{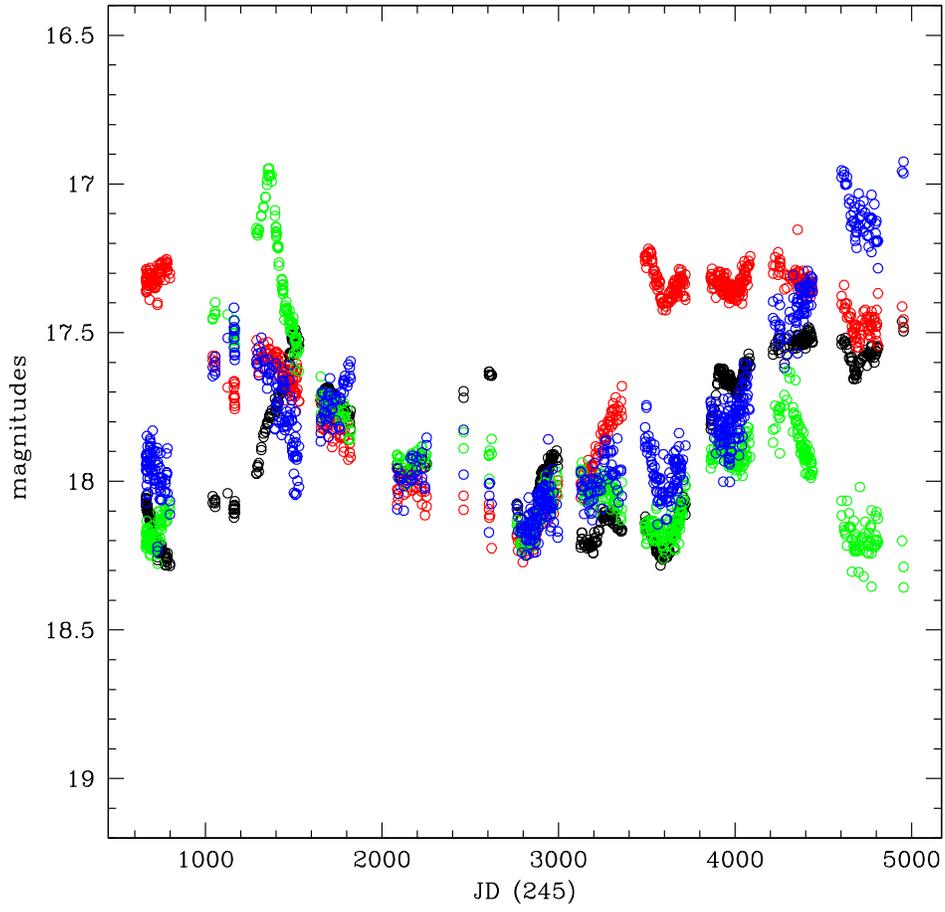}
\caption{OGLE light curves for components A (black), B (red), C (green) and D (blue) arbitrarily displaced in magnitudes to match around JD 2900. \label{ogle}}
\end{figure}

\begin{figure}[h]
%\plotone{/home/emg/2237/ogle/dif.eps}
\plotone{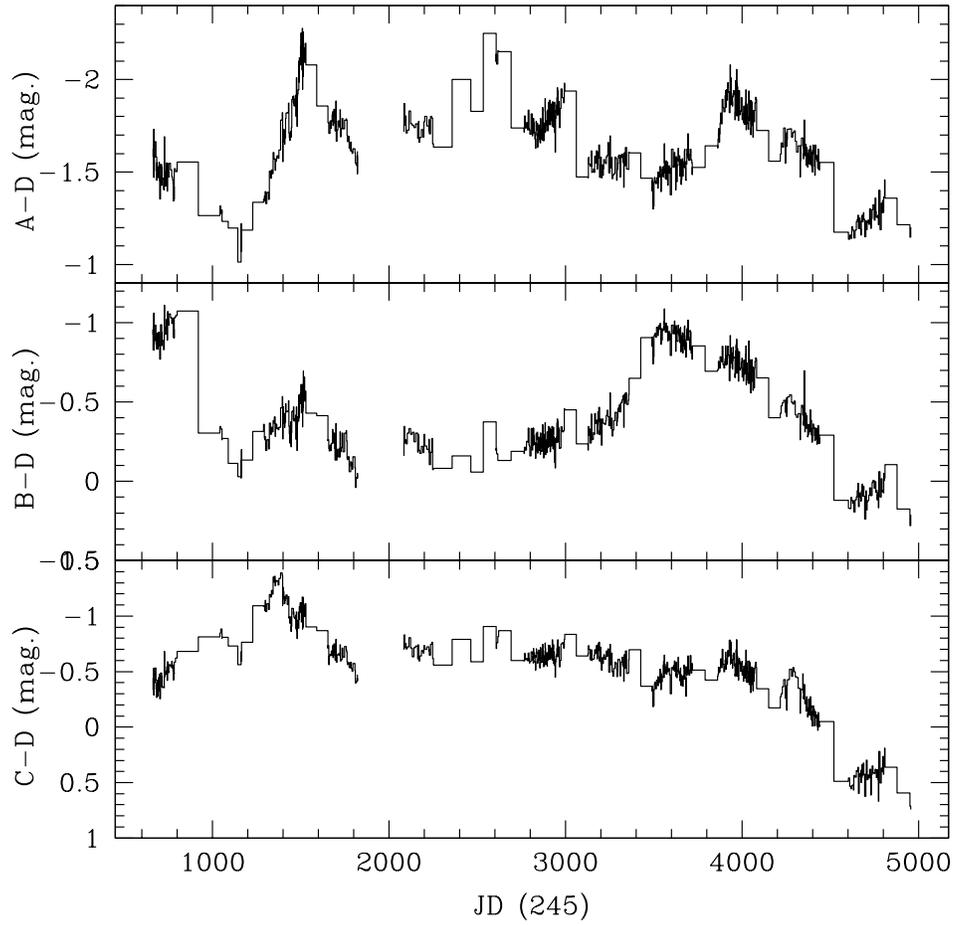}
\caption{A-D, B-D and C-D light curves from OGLE monitoring of Q~2237+0305.  \label{dif}}
\end{figure}

As commented in \S 1, lacking on regularly sampled X-Ray monitoring of any gravitational lens system, the best data available to explore the application of the caustic crossing statistic method are the optical light curves of Q 2237+0305 (Wozniak et al. 2000, Udalski et al. 2006). In principle, the main drawback to use optical monitoring is that after convolution with typical source sizes ($\sim4^{+2.4}_{-3.1}\sqrt{m/0.3 M_\odot}$ light-days according to Jim\'enez-Vicente et al. 2012) single caustic crossings may be mistaken with other type of events\footnote{Note, however, that the freedom of choice between alternative explanations will be significantly constrained by the comparison between several events and by the knowledge of the transverse velocity.} (cusp or multiple caustic crossing, caustic or cusp touching, etc.; see, e.g., Kochanek 2004, Mosquera et al. 2009). Our estimates for the average track length to obtain one caustic crossing (see Eq. \ref{track}), are in the ranges: 19$\pm 1$  to 33$\pm 2$  (images A and B), 22$\pm 1$  to 38$\pm 2$  (image C), and 12$\pm 1$ to 21$\pm 1$  (image D) light-days when the average mass of the PDMF, $\langle m \rangle$, ranges from 0.08 to 0.30$M_\odot$. Thus, the typical size is small enough as to allow the isolated detection of caustic crossings, but caustic blending is also possible within the range of uncertainties. Bearing in mind this limitation, we are going to estimate the number of caustic crossings in the OGLE optical light curves for Q 2237+0305 that extend for more than 12 years (Wozniak et al. 2000, Udalski et al. 2006). In Figure \ref{ogle} we represent OGLE light curves displaced to match each other in the interval between JD (245) 1500 and JD 3000. Except for epochs greater than JD 4500, the light curve of image D seems to always match the light curves of two or three of the other images, not showing any conspicuous feature (uncorrelated with the other light curves) that can be an evidence of strong microlensing. In fact, to our knowledge, no microlensing event has been reported for this image in the literature. For these reasons we will use the A-D, B-D, and C-D light curves (Figure \ref{dif}) to show possible caustic crossings in A, B, and C after removing the intrinsic source variability (the time delay between images is very small).

Several caustic crossing events have been reported in the literature for image A. There is a clear high magnification event at JD 1500 that has been interpreted as a caustic crossing by Wyithe et al. (2000c), Shalyapin (2001), Goicoechea et al. (2003), Gil-Merino et al. (2006) and Abolmasov \& Shakura (2012). Kochanek (2004) found that both, a single caustic crossing or a double caustic crossing in a cusp, are acceptable explanations for this event. Mosquera et al. (2009) have proposed another event,  likely peaking at the gap at JD 3000. They interpreted this event as a double caustic crossing or a tangential caustic crossing. A third high magnification event at JD 4000 was studied by Eigenbrod et al. (2008a,b).  It is also interesting to note, as another possible caustic crossing candidate, the high peak at JD 2600 located in an undersampled region of the light curve.

Image B light curve presents the highest magnification event (about 1 magnitude) at JD 3600 interpreted by Eigenbrod et al. (2008a,b) as a caustic crossing. Combining OGLE and X-Ray data, Mosquera et al. (2013) also support the caustic crossing interpretation.  However, this event is  very broad as compared with the others present in the light curves and could be related to a phenomenology involving more than a single caustic crossing.  In the light curve of image B we also found an event centered at JD 1500 not previously identified by others but with shape and FWHM similar to that of the events at JD 1500 and JD 4000 in image A and of the event at JD 1360 in image C (see below). Finally, there is a 1 magnitude rising before JD 1000 in an undersampled region of the light curve that could be also related to caustic crossing.

In the light curve of image C there is a peak at JD 1360 that has been related either to a single caustic crossing by Shalyapin (2001), Anguita et al (2008) and Abolmasov \& Shakura (2012) or to a more complex phenomenology involving a cusp by Wyithe et al. (2000c) and Kochanek (2004). Wyithe et al. (2000c) also point out that a caustic could have been lost at the JD 1000 gap.

To estimate a lower limit for the peculiar velocity of the lens galaxy we will start considering, conservatively, only the events related in the literature to the caustic crossing phenomenology plus the event centered at JD 1500 in B, and counting only a caustic crossing per event irrespective of other possible interpretations\footnote{Although all these events (except JD 1500 in B) have been published as probable caustic crossings, alternative explanations are acceptable in most cases.}. This leads to 3 probable caustic crossings for image A, 2 for image B, 1 for image C, and 0 for image D. From the joint probability distribution for A, B, C and D and marginalizing in $\sqrt{m}$ we obtain a  lower limit of the galaxy peculiar velocity of $v_{eff} \geq 240\rm\,km\, s^{-1}$ at 68\% of confidence. Taking into account the range of marginalization in $\sqrt{m}$ and the $v_{eff} \propto \sqrt{\langle m \rangle}$ degeneracy, we can make explicit the  dependence with the stellar mass function by writing\footnote{{Although $\sqrt{\langle m \rangle}$ is the variable typically used in previous studies of caustic crossing, it can be formally better to use $\langle\sqrt{ m }\rangle$. In this case, the factor $\sqrt{\langle m \rangle/0.17M_\odot}$ can be conveniently restated as ${\langle m \rangle/\langle \sqrt{m} \rangle\over \sqrt{0.22 M_\odot} }={\mu_1/\mu_{1/2}\over \sqrt{0.22 M_\odot} }$, where $\mu_1$ and $\mu_{1/2}$ are the corresponding moments of the mass distribution.}},  $v_{eff}\gtrsim 240\sqrt{\langle m \rangle/0.17M_\odot}\rm\,km\, s^{-1}$.

\begin{figure}[h]
%\plotone{/home/emg/2237/ogle/eventos4_new.eps}
\plotone{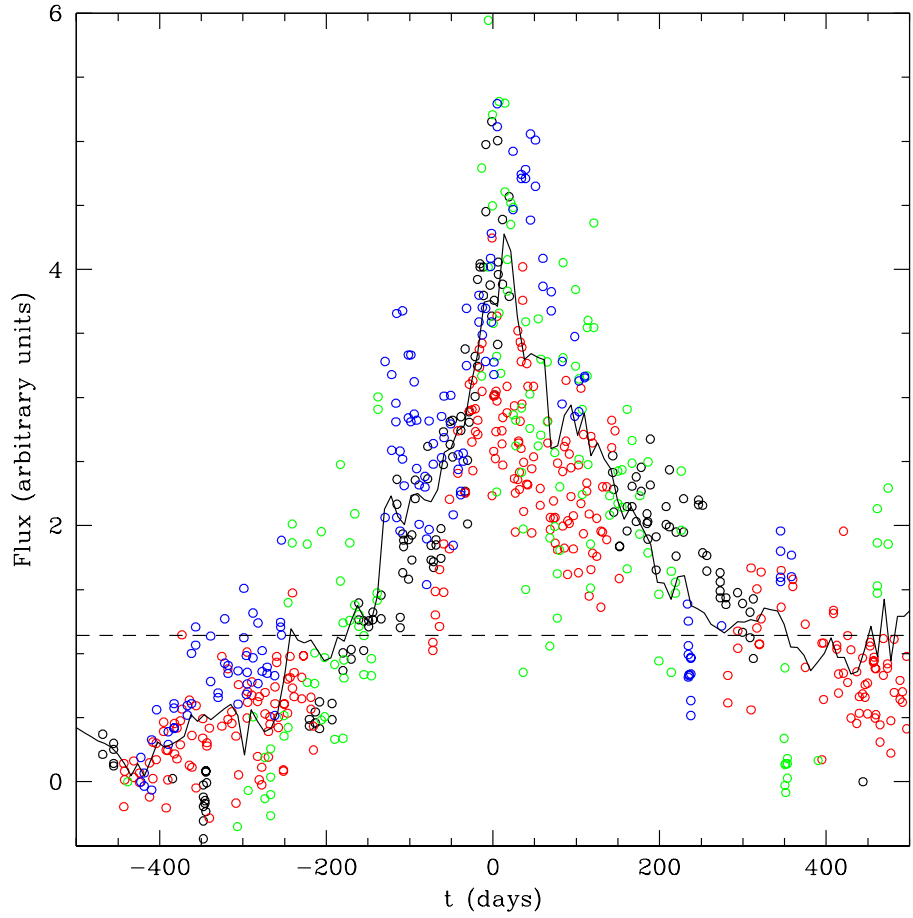}
\caption{Comparison of four HME: A/JD 1500 (black circles), A/JD 4000 (red circles), B/JD 1500 (green circles) and C/JD 1360 (blue circles) after subtracting a background level and normalizing. The continuous line is the average of the four events and the discontinuous matchs the profile inside the caustic.  \label{eventos4_new}}
\end{figure}

This lower limit is based on the events reported in the literature that, in most cases, have been  studied separately, using a small fraction of the light curve around each isolated event. Now, we have a global perspective of the four light-curves and we can compare the four events which are candidates to single caustic crossing: A/JD 1500, A/JD 4000, B/JD 1500, and C/JD 1360. To do this we need, taking as reference the caustic crossing approximation (see, e.g. Mosquera et al. 2009),  to (i) orient the events according to the sense of the caustic crossing (outside to inside), (ii) define and subtract the background level outside of the caustic and (iii) multiply the resulting profiles by an arbitrary factor to try to match them. The resulting comparison is plotted in Figure \ref{eventos4_new}. The resemblance between the light curves of  the four events is very noticeable. The average profile (see Figure \ref{eventos4_new}) looks more pointed than a Gaussian and asymmetric. Although the gaps in the sampling and the uncertainties in the definition of the background levels induce, indeed, some allowances in the comparison, the good matching and the shape of the average profile support the identification as single caustic crossings of the four events. The size, $r_s=FWHM/2.35$, of the average profile of the four events is roughly about 160 days. This size is substantially smaller than the typical separation between caustics observed in the light curves ($\gtrsim 1000$ days in A, $\gtrsim 2000$ days in B, and $\gtrsim 3000$ days in C) supporting the consistence of all the reasoning. For a velocity of 240$\rm\,km\, s^{-1}$, the estimated time lapse of 160 days results, after marginalizing in $\sqrt{\langle m \rangle}$, in a size of $r_s=1.3$ light days that can be interpreted as a lower limit to the radius of the source ($r_s\gtrsim 1.3 \sqrt{\langle m \rangle/0.17M_\odot}$, taking into account the $v_{eff} \propto \sqrt{\langle m \rangle}$ degeneracy). Corrections for unnoticed caustic blending will push the velocity upwards and, hence, will not affect to these two lower limits. On the contrary, caustic touching and passing nearby are two types of events that, counted as caustic crossings, may invalidate the lower limit. However, these events are  expected to be broader, smoother, and have a variety of shapes and FWHM. The resemblance between the light curve profiles of the four events do not support this possibility.

To obtain an estimate of the velocity we should include other possible events: (i) the high magnification peak at JD 2600 in A, (ii) a second caustic crossing in the event at JD 3000 in A, and, (iii) a second caustic crossing in the broad event at JD 3600 in B. This leads to 5, 3, 1, and 0 caustic crossings for images A, B, C and D respectively. The expected velocity from the marginalized (in $\sqrt{\langle m \rangle}$) joint probability distribution of the four images is $v_{eff}=493\pm246\rm\,km\, s^{-1}$ and the corresponding source radius $r_s=2.7\pm 1.3$ light-days ($v_{eff}\simeq(493\pm246)\sqrt{\langle m \rangle/0.17M_\odot}\rm\,km\, s^{-1}$ and $r_s\simeq(2.7\pm 1.3)\sqrt{\langle m \rangle/0.17M_\odot}$ light-days, taking into account the $v_{eff} \propto \sqrt{\langle m \rangle}$ degeneracy). In principle these can be reasonable estimates of the velocity and the size. However, although we have tried to include now all the events that can be reasonably interpreted in terms of caustic crossings, notice that: (i) some caustic crossings may be lost in the gaps at JD 1000 and JD 2000 {(correcting the observation time for the gaps, the velocity and the size will increase a $\sim$10\%)}, (ii) a third caustic may be involved in the broad event at JD 3600 in B, and (iii) caustic crossings resulting in relatively small amplitude events may be hidden. Finally, notice also that for epochs greater than JD 4500 the light curve of image D seems to start a steep rising. 

Our $v_{eff}$ estimate is in good agreement with the upper limits of 500$\rm\,km\, s^{-1}$ obtained by Wyithe et al. (1999) and of $685\rm\,km\, s^{-1}$ derived by Gil-Merino et al. (2005) and with the lower limit of 338$\rm\,km\, s^{-1}$ inferred by Poindexter \& Kochanek (2010).  On the other hand, our estimate of the half light radius, $R_{1/2}=1.18r_s=4.2\pm 2$ light-days (taking $M=0.3 M_\odot$), is also in agreement within errors with the values obtained by Eigenbrod et al. (2008b; $R_{1/2}=3.0\pm 2$ light-days with velocity prior $v_{eff}=410\rm\,km\, s^{-1}$), Sluse et al. (2011; $R_{1/2}=3.4^{+6.4}_{-2.4}$ light-days with velocity prior $v_{eff}=685\rm\,km\, s^{-1}$), Poindexter \& Kochanek (2010; $R_{1/2}=5.4\pm 3.2$ light-days) and  Mosquera et al. (2013; $R_{1/2}=9.9^{+5.1}_{-3.3}$ light-days).

We have estimated the effective transverse velocity, $v_{eff}$, that results from the composition of the velocity of the lens galaxy with the movement of the caustics caused by the random stellar kinematics. According to the simulations by Kundic \& Wambsganss (1993), the two effects are about equal when we compare the rms velocity of the stars in the plane of the galaxy perpendicular to the line of sight,  $\sqrt{2} \sigma_*$ ($\sigma_*$ is the one dimensional stellar velocity dispersion), and the transverse velocity of the galaxy, $v_t$,

\begin{equation}
v_{eff}=\sqrt{v_t^2+2 \sigma_*^2},
\label{KW}
\end{equation}
For Q~2237+0305 the one dimensional velocity dispersion is $\sigma_*=172 \rm\, km\, s^{-1}$ (Trott et al. 2010) and the random velocity of caustics is $\sqrt{2} \sigma_*=243 \rm\, km\, s^{-1}$ comparable to our lower limit for the effective velocity\footnote{Following Kundic \& Wambsganss (1993) we have adopted in Equation \ref{KW} a $\sqrt{2}$ effectiveness factor relating the transverse and random velocities but this factor could depend on optical depth and shear (Wyithe et al. 2000a).}. This makes statistically consistent the caustic identifications made for obtaining the lower limit for $v_{eff}$. Applying more restrictive criteria, we would have counted less caustics than expected from the effect of the random motion of the stars alone.

On the other hand, from the estimate of  $v_{eff}$ we obtain,

\begin{equation}
v_t =\sqrt{v_{eff}^2-2 \sigma_*^2} \simeq(429\pm246)\sqrt{\langle m \rangle/0.17M_\odot}\rm\,km\, s^{-1},
\end{equation}
for the transverse velocity of the lens galaxy. If we take, $\langle v_t\rangle=\sqrt{2} \sigma_{pec}$ where $\sigma_{pec}$ is the one dimensional velocity dispersion of the distribution of peculiar velocities of the galaxies, we finally obtain, $\sigma_{pec} \simeq(303\pm174)\sqrt{\langle m \rangle/0.17M_\odot}\rm\,km\, s^{-1}$. {Notice that, according to Poindexter \& Kochanek (2010), the  effective velocity is dominated by the lens velocity and that streaming velocities such as our motion relative to the CMB amount $\lesssim$10\% of the peculiar velocity.}

The lower limits and estimates obtained with the optical light curves are consistent and encouraging, but X-Ray observations are needed to safely identify all the caustics.  Chen et al. (2011) present X-Ray monitoring of Q~2237+0305 covering a long period of time. Mosquera et al. (2013) analyzed these data finding a reasonable agreement with the optical light curves. Unfortunately, the sampling is sparse and irregular. According to the $\sim$1 light-day size of the (hard) X-ray source inferred by Mosquera et al. (2013) and using $v_{eff}=493\rm\,km\, s^{-1}$  a caustic crossing event will last $\sim 61\rm\, days$. Thus, a regular monitoring with a sampling significantly better than two months will be needed to find out all the caustics.

\section{Conclusions}

We have discussed the application of the statistics of caustic crossings to the study of quasar microlensing, an effect that provides unique information about the physical parameters of lensed quasars and lens galaxies.  We have studied three of these parameters, the peculiar velocity of the lens galaxy, the slope of the lens galaxy stellar mass function (specifically the lower mass end slope of the PDMF, $\alpha_1$) and the quasar source size. We have obtained the following results:

1 - The effect of changing the shape of the PDMF, even in the wide range of lower mass end slopes used in the simulations (that change drastically the relative abundance between low and high mass stars), is unable to break the  $v_{eff} \propto \sqrt{\langle m \rangle}$ degeneracy except at a very low level of significance in $\langle m \rangle$.  Thus, we find a dependence of the average number of caustic crossings with the average mass,  $\langle n \rangle  \propto {v_{eff} /  \sqrt{\langle m \rangle}}$, equivalent to the theoretical relationship for the case of identical mass particles. The fact that this relationship holds for strong changes in the shape of the PDMF (induced in our calculations by the changes in $\langle m \rangle$) make us think that this relationship is rather generic.

2 - Our simulations show that, in what respects to the PDMF, the average number of caustic crossings $\langle n \rangle$ is sensitive only to the average mass of the stellar distribution, $\langle m \rangle$. This extends the result previously found by others considering Salpeter law with a lower mass cutoff at $0.1M_\odot$ (see Witt, Kayser \& Refsdal, 1993; Wyithe et al. 2000a), to the case of mass functions with a great abundance of low mass stars. In addition, for realistic values of $\langle m \rangle$, the average number of caustic crossings, $\langle n \rangle$, is much more sensitive to $v_{eff}$ than to the mean mass of the PDMF. This limits the usefulness of the causting crossing rate to study the PDMF but opens the possibility of determining $v_{eff}$ and the quasar source size, $r_s$.

3 - We have applied the statistics of caustic crossings to the best data available, the Q 2237+0305 optical light curves from OGLE. Accepting as good the caustic crossing identifications published in the literature, plus one additional proposed by us, we infer lower limits for the effective velocity and the size of the source of $v_{eff}\gtrsim 240\sqrt{\langle m \rangle/0.17M_\odot}\rm\,km\, s^{-1}$ and $r_s \gtrsim 1.4 \sqrt{\langle m \rangle/0.17M_\odot}$ at 68\% of confidence. 

4 - Tentative identification of 3 additional caustic crossing events leads to estimates of $v_{eff}\simeq (493\pm 246)\sqrt{\langle m \rangle/0.17M_\odot}\rm\,km\, s^{-1}$ for the effective velocity and of $r_s \simeq (2.7\pm 1.3)\sqrt{\langle m \rangle/0.17M_\odot}$ light-days for the source size. This size is substantially smaller than the average separation between caustics supporting the consistence of the calculations. From $v_{eff}$ and the stellar velocity dispersion in Q~2237+0305 we estimate a transverse peculiar velocity of  $v_t  \simeq(429\pm246)\sqrt{\langle m \rangle/0.17M_\odot}\rm\,km\, s^{-1}$ for the lens galaxy.

\acknowledgements

We thank C.S. Kochanek for helpful discussions, suggestions and comments. We
are also grateful to the anonymous referee for suggestions that helped to improve the
manuscript. JJV is supported by the Spanish Ministerio de Econom\'{\i}a y Competitividad through grant AYA2011-24728 and by the Junta de Andaluc\'{\i}a through project FQM-108. EM, TM, OA and
JAM were supported by the Spanish MINECO with the grants AYA2010-21741-C03-01 and AYA2010-21741-C03-02. JAM was also supported by the Generalitat Valenciana with the project PROMETEOII/2014/060.

%\begin{figure}[h]
%\plotone{mle_1d.eps}
%\caption{marginalized 1D MLE  \label{mle_1d}}
%\end{figure}

\end{document}